\documentclass[intlimits,twoside,a4paper]{article}

\usepackage[cp1251]{inputenc}

\usepackage[eqsecnum]{cmpj3}

\usepackage{bm}

\issue{2019}{22}{1}{13601}
\doinumber{10.5488/CMP.22.13601}

\title[Hydrodynamic correlations in isotropic fluids and liquid crystals]{Hydrodynamic correlations in isotropic fluids and liquid crystals 
       simulated by multi-particle collision dynamics}

\author{H. H\'{\i}jar}
\address{Engineering School, La Salle University Mexico, Benjamin Franklin 45, 
         06140, Mexico City, Mexico}

\date{Received November 14, 2018, in final form February 11, 2019}

\begin{document}

\maketitle

\begin{abstract}

Multi-particle collision dynamics is an appealing numerical technique aiming at 
simulating fluids at the mesoscopic scale. It considers molecular details in a 
coarse-grained fashion and reproduces hydrodynamic phenomena. Here, the 
implementation of multi-particle collision dynamics for isotropic fluids is 
analysed under the so-called Andersen-thermostatted scheme, a particular 
algorithm for systems in the canonical ensemble. This method gives rise to 
hydrodynamic fluctuations that spontaneously relax towards equilibrium. This 
relaxation process can be described by a linearized theory and used to calculate 
transport coefficients of the system. The extension of the algorithm for nematic 
liquid crystals is also considered. It is shown that thermal fluctuations in the 
average molecular orientation can be described by an extended linearized scheme. 
Flow fluctuations induce orientation fluctuations. However, orientational 
changes produce observable effects on velocity correlation functions only when
simulation parameters exceed their values from those used in previous 
applications of the method. Otherwise, the flow can be considered to be independent 
of the orientation field.

\pacs 61.20.Ja, 05.20.Jj, 05.40.-a, 83.80.Xz
\end{abstract}

\section{Introduction}
\label{introduction_section}

Simulation of complex fluids, e.g., chemically reacting systems, colloids, and 
liquid crystals (LCs), represents a major task in computational physics of 
condensed matter~\cite{001,002}. Such fluids are characterised by phenomena 
occurring at widely separated length and time-scales, all of which are relevant 
in determining the observed behaviour. The interest in complex systems usually 
focuses on the dynamics of some microscopic degrees of freedom interacting with a 
solvent. Although it is essential to reproduce the correct behaviour of the 
solvent over long distances and large times, its molecular details are 
irrelevant. Traditional simulation techniques, e.g., molecular dynamics (MD), are 
very successful in describing equilibrium states of atomistic 
ensembles~\cite{003}, though it is not feasible to use them to simulate the 
solvent due to the enormous number of degrees of freedom and the extremely large 
time ranges that must be covered~\cite{003a}.

To incorporate the collective effects, e.g., thermal fluctuations and flow, in 
simulations of liquids is a challenge that has motivated the development of 
novel computational approaches which take into account essential dynamical 
properties and allow for coupling with the interesting microscopic degrees of 
freedom, and yet are simple enough to be simulated for long times and distances.  
Stokesian dynamics~\cite{006} and the lattice Boltzmann method~\cite{007} are 
examples of such approaches. Multi-particle collision dynamics (MPCD) is probably 
the most successful of such mesoscopic algorithms. It was introduced by 
Malevantes and Kapral~\cite{001,002}, and simulates fluids by means of particles 
whose positions and velocities are considered as continuous variables. The 
microscopic details of these particles are not specified. Instead, their time 
evolution is treated in a simplified form through collective stochastic 
collisions which preserve momentum and energy. This property allows MPCD to 
recover, over long simulation times, the hydrodynamic equations of mass, 
momentum, and heat propagation. In addition, the stochastic character of MPCD 
produces fluctuations and Brownian forces~\cite{008,009}.

A considerable number of soft condensed matter systems have been successfully 
simulated under MPCD schemes. Recent examples include sediment-water interface 
flow~\cite{010}, bistable biochemical systems~\cite{011}, and two-dimensional 
one-component plasma~\cite{012}, to mention but a few. Furthermore, modified 
MPCD rules have been used to extend simulations towards complex solvents, e.g., 
viscoelastic fluids~\cite{013} and binary mixtures~\cite{014}. Very recently,
algorithms for simulating nematic liquid crystals (NLCs) using the principles of 
MPCD have been also proposed~\cite{015,016}. They allow one to reproduce 
hydrodynamic and elastic characteristics of such phases, still being 
potentially able to be coupled with microscopic degrees of freedom. These 
algorithms simulate isotropic-nematic phase transitions, consistent dynamics
for annihilation of defects, and molecular reorientation under flow.

Here, basic implementations of MPCD for simple liquids and NLCs are considered. 
Simulations for equilibrium states are presented, where systems are in contact 
with the thermal bath based on an Andersen thermostat that keeps them at a fixed 
temperature. In order to exhibit the ability of MPCD to reproduce hydrodynamic 
behaviour and to emphasise its stochastic character, attention is focused on the
analysis of the spectra of hydrodynamic fluctuations produced by the algorithm 
in both simple liquids and nematics. There is discussed a very good agreement 
that exists between correlation functions obtained from the numerical 
implementation with those derived from linearized hydrodynamic models of liquids 
and LCs. Following the MPCD model for nematics (MPCD-N) introduced by Shendruk 
and Yeomans in reference~\cite{016}, simulated particles are assumed to be 
slender rods and velocity gradients generate torques on them. On the other hand, 
the effect of reorientation on flow, commonly referred to as \emph{backflow}, is 
incorporated in the velocity update stage through a term that converts the 
angular momentum generated by reorientation into orbital angular momentum. 
Numerical experiments are conducted in favour of exploring the resulting effects 
of this orientation$\leftrightarrow$velocity coupling scheme on the relaxation 
of hydrodynamic fluctuations. It is found that the influence of flow on 
orientation is very well accounted for by the linearized form of Jeffery's shear 
alignment rule. By contrast, backflow does not produce appreciable effects on the 
spectra of hydrodynamic correlations as long as its associated simulation 
parameter is maintained within the range proposed in the original version of the 
method. Through these results it is shown that MPCD has the potential for 
simulating hydrodynamic phenomena at the mesoscopic scale when extended to 
mesomorphic phases. 

\section{MPCD algorithm for isotropic fluids}
\label{mpcd_algorithm_isotropic_section}

Multi-particle collision dynamics simulates fluids composed of point particles. 
In the simplest case, all these particles have the same mass, $m$. They move 
within a simulation box with sizes $L_{x}$, $L_{y}$, and $L_{z}$ in the $x$, 
$y$, and $z$ directions of a Cartesian coordinate system, respectively. In what 
follows, the total number of simulated particles will be denoted by $N$, 
whereas positions and velocities of particles will be grouped in the matrices 
$r_{i,\alpha}$ and $v_{i,\alpha}$, respectively, where indices $i$ and $\alpha$ 
indicate, respectively, particle number and Cartesian coordinate. Thus,
$i = 1, 2, 3, \ldots, N$; and $\alpha = x,y,z$. Quantities $r_{i,\alpha}$ and 
$v_{i,\alpha}$ are considered to be continuous functions of the simulation time, 
$t$. Multi-particle collision dynamics promotes the evolution of $r_{i,\alpha}$ 
and $v_{i,\alpha}$ through a succession of streaming and collision events taking 
place at regular time intervals of size $\Delta t$. 
The precise steps of the MPCD algorithm are described below.

\subsection{Streaming}
\label{streaming_section}

For simulations of homogeneous systems, streaming consists of a simple ballistic 
displacement of the particles during the time-step $\Delta t$. More precisely, 
given the current state variables, $r_{i,\alpha}(t)$ and $v_{i,\alpha}(t)$, 
positions at next time-step are given by
\begin{equation}
r_{i,\alpha}(t+\Delta t) = r_{i,\alpha}(t) + v_{i,\alpha}(t) \Delta t. 
\label{streaming}
\end{equation}

It is worth noting that streaming must be followed by a rule for handling the 
escape of particles through the boundaries of the simulation box. To 
approximate the behaviour of macroscopic systems in thermodynamic equilibrium, 
typical periodic boundary conditions (PBCs) are preferred~\cite{003}. 

\subsection{Collision}
\label{collision_section}

The physical aim of the collision step is to promote the momentum exchange 
between particles. 
Collisions are performed by imagining the simulation box as composed of 
equal-sized cubic cells. For this purpose, it is considered that $L_{x}$, 
$L_{y}$, and $L_{z}$, are multiples of a unit length, $a$, i.e., $L_{x}=n_{x}a$, 
$L_{y} = n_{y} a$, and $L_{z} = n_{z}a$, with $n_{x}$, $n_{y}$, and $n_{z}$ 
integers. Particles located within the same cell collide. Therefore, the 
imaginary cells are referred to as collision cells. The number of particles 
within collision cells could be different and vary as function of time because 
particles might come in and out from them.

Here, for concreteness, discussion will be focused on the MPCD method based on 
the application of an Andersen thermostat (MPCD-AT), where relative velocities 
within collision cells are replaced by new velocities sampled from the 
Maxwell-Boltzmann distribution at temperature $T$. 
Two possible versions of MPCD-AT exist differing from one another by their 
capacity to preserve the angular momentum. For simulations without angular 
momentum conservation (MPCD-AT$-$a), velocities are updated according to
\begin{equation}
v_{i,\alpha} (t+\Delta t) = \bar{v}^{\,\text{c}}_{\alpha}(t)
			  + \xi_{i,\alpha} - \bar{\xi}^{\,\text{c}}_{\alpha}\,,
\label{collision_ma}
\end{equation}
where $\bar{v}^{\,\text{c}}_{\alpha}(t)$ is the centre of mass velocity of the 
cell containing the $i$-th particle, $\xi_{i,\alpha}$ is the assigned stochastic 
velocity, and 
$\bar{\xi}^{\,\text{c}}_{\alpha} = \sum_{j\in c} \xi_{j,\alpha}/N^{\text{c}}$. In the 
last definition, summation extends over all the $N^{\text{c}}$ particles containing the 
cell where $i$ is located. Notice that hereafter a superscript ``$\text{c}$'' will 
be used to indicate properties measured at MPCD collision cells, and should not 
be confused with a variable that can take numerical values. It can be easily 
verified that equation~(\ref{collision_ma}) preserves linear momentum after 
collision. Nevertheless, it generates the change in angular momentum,
\begin{equation}
\Delta L^{\text{c}}_{\alpha} 
= \sum_{j \in c} \varepsilon_{\alpha \alpha^{\prime} \beta^{\prime}}
 \left(r_{j,\alpha^{\prime}}-\bar{r}^{\,\text{c}}_{\alpha^{\prime}}\right)
 \left(v_{j,\beta^{\prime}}-\xi_{j,\beta^{\prime}}\right),
\label{ang_momentum_change_001}
\end{equation}
where $\bar{r}^{\,\text{c}}_{\alpha}$ is the centre of mass position in the cell, 
$\varepsilon_{\alpha\alpha^{\prime}\beta^{\prime}}$ is the Levi-Civita 
symbol, and the summation over Greek repeated indices is implied, a convention that will 
be followed from now on, unless the contrary is explicitly indicated. 
Simulations with conservation of local angular momentum, MPCD-AT$+$a, are 
accomplished by subtracting from equation~(\ref{collision_ma}) the amount of 
angular momentum arising from stochastic velocity sampling, namely~\cite{021}
\begin{equation}
v_{i,\alpha}(t + \Delta t) = \bar{v}^{\,\text{c}}_{\alpha}(t) + \xi_{i,\alpha} 
                           - \bar{\xi}^{\,\text{c}}_{\alpha} 
			   -\varepsilon_{\alpha \alpha^{\prime} \beta^{\prime}}
			    J^{\text{c}}_{\alpha^{\prime}\alpha^{\prime\prime}} (t) 
                            \Delta L^{\,\text{c}}_{\alpha^{\prime\prime}}(t)
			    \left( r_{i,\beta^{\prime}} - \bar{r}^{\,\text{c}}_{\beta^{\prime}}
                            \right),
\label{collision_pa}
\end{equation}
where the last term on the right-hand side is responsible for angular 
momentum conservation. There, $J^\text{c}_{\alpha\beta}$ is the inverse of the 
local moment of inertia tensor.

In both MPCD-AT$-$a and MPCD-AT$+$a, temperature is automatically controlled by 
the stochastic velocity sampling, so simulations proceed in the canonical 
ensemble.

\subsection{Grid shift}
\label{grid_shift_section}

Implementations of MPCD based solely on streaming and collision are not Galilean 
invariant, an effect that becomes more notorious in simulations conducted at 
small mean free paths, $l_{\textrm{m}} = \Delta t(k_{\textrm{B}}T/m)^{1/2}$, 
where $k_{\textrm{B}}$ is the Boltzmann constant. The reason is that for small 
$l_{\textrm{m}}$, particles collide repeatedly with those in their 
neighbourhood and become correlated over long time periods. In order to restore 
Galilean invariance, Ihle and Kroll proposed to perform a stochastic 
displacement of the grid of collision cells before the momentum exchange 
event~\cite{022}. It is usual to conduct the grid shift independently along each 
Cartesian axis, using uniform distributions within the range 
$\left[-a/2,a/2\right]$. The stochastic grid shift helps particles to exchange 
momentum with a broader set of neighbours and allows the system to recover 
molecular chaos. In practice, it is easier to program this step by displacing 
all the particles in the system by the same random vector, applying the 
corresponding boundary conditions for those particles leaving the simulation 
box. Once particles have been accommodated, collision takes place. Then, 
particles are moved back by reversing the stochastic displacement and applying 
the proper boundary conditions.

\section{Hydrodynamic fluctuations in MPCD}
\label{hydrodynamic_fluctuations_mpcd_section}

\subsection{Linearized hydrodynamics}
\label{linearized_hydrodynamics_section}

Steps in MPCD are designed to satisfy the mass and momentum conservation. Malevanets 
and Kapral~\cite{001,002} showed that when these steps are applied over long simulation times, 
they permit the system to solve the Navier-Stokes equations. 
Moreover, MPCD is intrinsically stochastic, therefore density and velocity 
fields suffer from small fluctuations at the local level. Analysing the 
behaviour of fluctuations produced by the 
implementation is fundamental for diverse reasons. First, it is important to 
verify that spontaneous changes in hydrodynamic fields are consistent with a 
hydrodynamic description, since such changes will be responsible for producing 
Brownian forces on solute particles~\cite{023}. Second, by analysing the spatial 
and temporal decay of fluctuations it is possible to measure the properties of the 
simulated systems, namely, viscosity, diffusion, and sound attenuation 
coefficients~\cite{008}. 
The results of such measurements can be compared against predictions of 
theoretical treatments of the algorithm to reinforce the understanding of the 
method and give reliance in using it in more complex situations.

Since MPCD-AT$+$a and MPCD-AT$-$a are intrinsically thermostatted, fluids 
simulated by them require the specification of two fields only. They are the 
density and velocity fields, $\rho(\vec{r},t)$ and $V_{\alpha}(\vec{r},t)$,
respectively, where $\vec{r}$ is the position vector. These fields obey the 
continuity and momentum conservation equations,
\begin{equation}
\frac{\partial \rho}{\partial t} 
= -\frac{\partial}{\partial x_{\alpha}} \left( \rho V_{\alpha}\right) ,
\label{continuity}
\end{equation}
\begin{equation}
\rho \left( 
     \frac{\partial }{\partial t} 
     + V_{\beta} \frac{\partial}{\partial x_{\beta}} 
     \right) 
     V_{\alpha} 
= - \frac{\partial}{\partial x_{\beta}} \sigma_{\alpha \beta}\,,
\label{navier_stokes}
\end{equation}
respectively, where $\sigma_{\alpha \beta} = \sigma_{\alpha \beta} (\vec{r},t)$ 
is the stress tensor. 
By introducing the thermodynamic pressure, $p(\vec{r},t)$, and assuming a linear 
dependence of the viscous stress on the velocity gradient 
$\partial V_{\alpha}/\partial x_{\beta}$, $\sigma_{\alpha \beta}$ can 
be written in the form
\begin{equation}
\sigma_{\alpha \beta} = -p \delta_{\alpha \beta} 
                      + \eta_{\alpha \beta \alpha^{\prime} \beta^{\prime}} 
                        \frac{\partial V_{\beta^{\prime}}}
                             {\partial x_{\alpha^{\prime}}}\, ,
\label{stress_tensor}
\end{equation}
where $\eta_{\alpha \beta \alpha^{\prime} \beta^{\prime}}$ is the viscous 
tensor, whose particular form depends on the employed MPCD algorithm. Explicit 
expressions for viscous contributions in MPCD-AT$\pm$a have been derived, e.g., 
in reference~\cite{023a}. In case of no angular momentum conservation,  
$\sigma_{\alpha \beta}$, reads as~\cite{023b}
\begin{equation}
\sigma_{\alpha \beta} = -p \delta_{\alpha \beta} 
		      + \bar{\eta}^{\,\text{kin}}
                      \left( \frac{\partial V_{\beta}}{\partial x_{\alpha}}
                            +\frac{\partial V_{\alpha}}{\partial x_{\beta}}
                            -\frac{2}{3} \delta_{\alpha \beta}
			     \frac{\partial V_{\alpha^{\prime}}}{\partial x_{\alpha^{\prime}}}
                      \right)
		      +\bar{\eta}^{\,\text{col}} 
                       \frac{\partial V_{\alpha}}{\partial x_{\beta}}\,,
\label{stress_tensor_ma}
\end{equation}
where viscosity coefficients $\bar{\eta}^{\,\text{kin}}$ and $\bar{\eta}^{\,\text{col}}$ 
represent contributions due to kinetic and collision processes, respectively. 
The former comes from the transverse momentum transport produced by motion of 
particles and is the dominant contribution to the viscosity of gas phases. 
Collisional viscosity is related with the redistribution of momentum caused by 
multi-particle collision events. For MPCD-AT$-$a, 
$\bar{\eta}^{\,\text{kin}}$ and $\bar{\eta}^{\,\text{col}}$ explicitly read 
as~\cite{023b}
\begin{equation}
	\bar{\eta}^{\,\text{kin}} = \frac{\bar{N}^{\textrm{c}} k_{\textrm{B}} T \Delta t}{a^{3}} 
		    \left( \frac{\bar{N}^{\textrm{c}}}{\bar{N}^{\textrm{c}}-1 + \re^{-\bar{N}^{\textrm{c}}}}
		          -\frac{1}{2}
		    \right),
\label{viscosity_kin_ma}
\end{equation}
\begin{equation}
	\bar{\eta}^{\,\text{col}} = \frac{m}{12 a \Delta t}\left(\bar{N}^{\textrm{c}}-1 + \re^{-\bar{N}^{\textrm{c}}}\right) ,
\label{viscosity_col_ma}
\end{equation}
where $\bar{N}^{\textrm{c}}$ represents the average numerical density at collision cells.

For MPCD-AT$+$a, the stress tensor can be written in the form~\cite{023b} 
\begin{equation}
\sigma_{\alpha\beta} = -p \delta_{\alpha \beta} 
                      +\eta
                      \left( \frac{\partial V_{\beta}}{\partial x_{\alpha}}
                            +\frac{\partial V_{\alpha}}{\partial x_{\beta}}
                            -\frac{2}{3} \delta_{\alpha \beta}
			     \frac{\partial V_{\alpha^{\prime}}}{\partial x_{\alpha^{\prime}}}
                      \right)
		      +\eta^{\text{V}} \delta_{\alpha \beta} 
		       \frac{\partial V_{\alpha^{\prime}}}{\partial x_{\alpha^{\prime}}}\,,
\label{stress_tensor_pa}
\end{equation}
where, $\eta = \eta^{\text{kin}} + \eta^{\text{col}}$ with
\begin{equation}
\eta^{\text{kin}} = \frac{k_{\textrm{B}} T \bar{N}^{\textrm{c}} \Delta t}{a^{3}}
		    \left(\frac{\bar{N}^{\textrm{c}}}{\bar{N}^{\textrm{c}} - \frac{5}{4}} -\frac{1}{2}
                    \right),
\label{viscosity_kin_pa}
\end{equation}
\begin{equation}
\eta^{\text{col}} = \frac{m \bar{N}^{\textrm{c}}}{24 a \Delta t}
		    \left( 1 - \frac{7}{5 \bar{N}^{\textrm{c}}}
                    \right),
\label{viscosity_col_pa}
\end{equation}
while $\eta^{\text{V}}$, i.e., the bulk viscosity coefficient, reduces to 
$\eta^{\text{V}} = \bar{\eta}^{\,\text{col}}/3$.

The system of hydrodynamic equations for MPCD-AT$\pm$a can be closed through the 
thermal equation of state, $p(r,t) = (T,\rho(\vec{r},t))$, a relation that for 
MPCD fluids has been proven to be of the ideal gas type~\cite{023}. 
Density and velocity fluctuations, $\delta \rho(\vec{r},t)$ and 
$\delta V_{\alpha}(\vec{r},t)$, are defined through
\begin{equation}
\rho(\vec{r},t) = \rho^{\textrm{eq}} + \delta \rho(\vec{r},t),
\label{density_fluctuation}
\end{equation}
and 
\begin{equation}
V_{\alpha}(\vec{r},t) = V^{\textrm{eq}}_{\alpha} + \delta V_{\alpha} (\vec{r},t),
\label{velocity_fluctuation}
\end{equation}
respectively, where $\rho^{\textrm{eq}}$ and $V^{\textrm{eq}}_{\alpha} = 0$, are 
the equilibrium density and flow fields.

In addition, it must be considered that due to the stochastic motion of MPCD 
particles, spontaneous stress may occur at the local level which is not 
originated by velocity gradients. Stochastic stress is considered to be an 
additive term to $\sigma_{\alpha \beta}$, $\Sigma_{\alpha \beta}$, which as 
customary is approximated as a Gaussian-Markov process (white noise). The 
fluctuation-dissipation theorem (FDT) relates two stochastic stress components 
at two different times and positions as
\begin{equation}
\left\langle 
             \Sigma_{\alpha\beta}\left(\vec{r}^{\,\prime},t^{\prime}\right) 
	     \Sigma_{\alpha^{\prime}\beta^{\prime}}\left(\vec{r},t\right) 
\right\rangle 
= 2 k_\text{B} T \eta_{\alpha \beta \alpha^{\prime}\beta^{\prime}} 
\delta\left(\vec{r}^{\,\prime}- \vec{r}\right)
\delta\left( t^{\prime} - t \right),
\label{stress_fdt}
\end{equation}
where $\delta\left(\vec{r}^{\,\prime}- \vec{r}\right)$ and 
$\delta\left( t^{\prime}-t \right)$ 
are Dirac delta functions in space and time domains, respectively.

Fluctuating fields are more conveniently studied in Fourier space. Hereafter, 
Fourier transforms are to be indicated by a tilde over the corresponding 
function. In Fourier domain, the FDT for stochastic stress reads as
\begin{equation}
\big\langle 
\tilde{\Sigma}_{\alpha\beta}\big(\vec{k}^{\,\prime},\omega^{\prime}\big) 
\tilde{\Sigma}_{\alpha^{\prime}\beta^{\prime}}\big(\vec{k},\omega\big) 
\big\rangle 
= 2 (2\piup)^{4} k_\text{B} T \eta_{\alpha \beta\alpha^{\prime}\beta^{\prime}} 
\delta\big(\vec{k}^{\,\prime}+\vec{k}\big)
\delta\left( \omega^{\prime}+\omega \right).
\label{stress_fdt_ft}
\end{equation}

By introducing unit vectors $\hat{e}_{1} = \vec{k}/k$, 
$\hat{e}_{2} = k \left[ \hat{e}_{z} - \left(\hat{e}_{1}\cdot \hat{e}_{z}\right) 
\hat{e}_{1}\right]/(k_{x}^{2}+k_{y}^{2})^{1/2}$, and 
$\hat{e}_{3} = \hat{e}_{1}\times \hat{e}_{2}$, and projections 
$\delta \tilde{V}_{\alpha} = \hat{e}_{\alpha} \cdot \delta\tilde{\vec{V}}$,
hydrodynamic fluctuations can be split into independent variables
$\delta \tilde{V}_{2} $, $\tilde{V}_{3}$, and 
$\left\{\delta\tilde{\rho}, \delta \tilde{V}_{1}\right\}$. 
Dynamic equations for these quantities are 
obtained after replacing equations~(\ref{density_fluctuation}) and 
(\ref{velocity_fluctuation}) into equations~(\ref{continuity}) 
and (\ref{navier_stokes}), retaining only linear terms in fluctuations, and  
incorporating a stochastic stress. Such equations explicitly read as
\begin{equation}
\left(-\ri \omega  + \nu k^{2}\right) \delta \tilde{V}_{\alpha}  
= \frac{\ri k}{\rho_{0}} \tilde{\Sigma}_{1\alpha}
\label{transverse_iso}
\end{equation}
for the transverse velocity components, $\alpha = 2,3$; and
\begin{equation}
\left(\begin{matrix}
	-\ri \omega                             & -\ri k \rho_{0}                      \\
	-\ri k c_{\textrm{T}}^{2} \rho_{0}^{-1} & -\ri \omega  + D_{\textrm{l}} k^{2} 
      \end{matrix}
\right) \cdot 
\left(\begin{matrix}
      \delta\tilde{\rho} \\
      \delta\tilde{V}_{1}
      \end{matrix}
\right) =
        \frac{\ri k}{\rho_{0}} 
\left(\begin{matrix}
      0 \\
      \tilde{\Sigma}_{11}
      \end{matrix}
\right)
\label{longitudinal_iso}
\end{equation}
for the so-called longitudinal fluctuations. In equations~(\ref{transverse_iso}) and 
(\ref{longitudinal_iso}), $\nu = (\eta^{\text{kin}}+\eta^{\text{col}})/\rho_{0}$ 
is the kinematic viscosity of the fluid; 
$c_{\textrm{T}} = [(\partial p/\partial\rho)_{T}]^{1/2} $ is the isothermal sound speed; 
and $D_{\textrm{l}}$ is the so-called longitudinal viscosity. For MPCD-AT$-$a 
$D_{\textrm{l}} = (4\eta^{\text{kin}}/3 + \eta^{\text{col}})/\rho_{0}$, while for 
MPCD-AT$+$a one has $D_{\textrm{l}} = (4\eta/3 + \eta^{\text{V}})/\rho_{0}$.

Autocorrelation functions (ACFs) between fluctuations in Fourier coordinates are 
defined as
\begin{equation}
C_{\rho}\big(\vec{k},\omega\big) = \big\langle 
                           \delta \tilde{\rho}    \big(\vec{k},\omega\big) 
                           \delta \tilde{\rho}^{*}\big(\vec{k},\omega\big) 
                           \big\rangle,
\label{density_acf_def}
\end{equation}
and
\begin{equation}
C_{\textrm{V}\alpha}\big(\vec{k},\omega\big) = \big\langle 
                              \delta \tilde{V}_{\alpha} \big(\vec{k},\omega\big) 
                              \delta \tilde{V}_{\alpha}^{*}\big(\vec{k},\omega\big) 
                              \big\rangle,
\label{velocity_acf_def}
\end{equation}
where no summation over repeated indices is implied. Explicit expressions for 
$C_{\rho}(\vec{k},\omega)$ and $C_{\text{V}\alpha}(\vec{k},\omega)$ can be obtained by 
solving equations~(\ref{transverse_iso})--(\ref{longitudinal_iso}) 
and using the FDT in Fourier space, equation~(\ref{stress_fdt_ft}). This 
procedure yields the classical expressions
\begin{equation}
C_{\rho}\big(\vec{k},\omega\big) 
= A \frac{\rho_{0} D_{\textrm{l}} k^{4}}{q^{2}(k) + D_{\textrm{l}}^{2}k^{4}}
\left\{
\frac{1-\omega/[2q(k)]}{[\omega-q(k)]^{2} + D_{\textrm{l}}^{2}k^{4}}
+\frac{1+\omega/[2q(k)]}{[\omega+q(k)]^{2} + D_{\textrm{l}}^{2}k^{4}}
\right\},
\label{density_acf}
\end{equation}
\begin{equation}
C_{\textrm{V1}}\big(\vec{k},\omega\big) 
= \frac{\omega^{2}}{\rho_{0}^{2}k^{2}} 
  C_{\rho}\big(\vec{k},\omega\big),
\label{velocity_long_acf}
\end{equation}
\begin{equation}
C_{\textrm{V2}}\big(\vec{k},\omega\big) 
= C_{\textrm{V3}}\big(\vec{k},\omega\big) 
= A \frac{1}{\rho_{0}}
    \frac{\nu k^{2}}{\omega^{2} + \nu^{2} k^{4}}.
\label{velocity_trans_acf}
\end{equation}

In equations~(\ref{density_acf})--(\ref{velocity_trans_acf}), $q(k)$ has been 
defined as $q(k) = c_{\textrm{T}} k [1-(D_{\textrm{l}}k/c)^{2}]^{1/2}$, and 
$A$ is a constant defined by 
$A = 2 (2 \piup)^{4} k_\text{B} T \delta(\vec{k} = 0) \delta(\omega = 0)$. 

It can be noticed that $C_{\rho}$ consist of two symmetric peaks, commonly 
referred to as the Brillouin peaks, centred at $\pm q(k)$. In MPCD-AT$\pm$a, 
the classical scattering spectrum lacks the central (Rayleigh) peak because the 
algorithm is thermostatted and no temperature fluctuations are sustained. 
In more general cases, temperature fluctuations in fluids cannot be neglected. 
They relax towards equilibrium following a diffusion type equation with an extra 
term that couples them with longitudinal velocity components~\cite{008}. 
Temperature fluctuations produce the previously mentioned Rayleigh peak and have 
a measurable effect on the propagating modes~\cite{021a}. In the presence of 
temperature fluctuations, Brillouin peaks are centered at the isentropic sound 
speed instead of being located at $c_{\textrm{T}}$. It is worth noting that 
some MPCD variations could be used to incorporate temperature fluctuations. 
With this aim, a natural alternative to use is stochastic rotation 
dynamics~\cite{001,002}, the first MPCD algorithm, because it incorporates 
mesoscopic heat flow. Stochastic rotation dynamics could be used at the 
momentum exchange step instead of the collision rule based on the Andersen 
thermostat and it can be even modulated to give conditions ranging from 
adiabatic to isothermal~\cite{008}.

It can be further observed that function $C_{\textrm{V1}}$ is also double 
peaked. However, due to the factor $\omega^{2}$ in front of it, $C_{\textrm{V1}}$ 
vanishes at $\omega = 0$. Correlation functions $C_{\textrm{V2}}$ and 
$C_{\textrm{V3}}$ consist of a single Lorentzian peak at $\omega = 0$. 

\subsection{Measurements}
\label{measurements_section}

In practice, measurements of ACFs are conducted by recording time-series of 
spatial Fourier transforms of density and velocity fluctuations, namely
\begin{equation}
\delta \bar{\rho} \big(\vec{k} , t_{l} \big) 
= \rho^{\textrm{eq}} \sum_{j=1}^{N} \exp\big[-\ri \vec{k}\cdot\vec{r}_{j}(t_{l}) \big], 
\label{time_series_density}
\end{equation}
and
\begin{equation}
\delta \bar{V}_{\alpha} \big(\vec{k} , t_{l} \big) 
= \sum_{j=1}^{N} v_{j,\alpha} \exp\big[-\ri \vec{k}\cdot\vec{r}_{j}(t_{l}) \big],
\label{time_series_velocity}
\end{equation}
where $t_{l}$ indicates the simulation time and the over-bar transformation over 
spatial domain only. Then, ACFs at two different times are calculated using the 
time-origin moving scheme, e.g.,
\begin{equation}
C_{\rho}\big(\vec{k}, t_{l+m} - t_{l}\big) 
= \frac{1}{N_{\text{s}} - m +1} 
  \sum_{l = 0}^{N_{\text{s}}-m} \delta \bar{\rho}\big(\vec{k}, t_{l+m} \big)   
                                \delta \bar{\rho}\big(\vec{k}, t_{l}   \big),
\label{correlations_time}
\end{equation}
where $N_{\text{s}}$ denotes the size of the recorded time-series. Finally, 
discrete Fourier transformation over time domain can be applied to get ACFs in
$\{\vec{k},\omega\}$ coordinates.

It should be stressed that due to the use of PBC, wave vector components are 
restricted to be multiples of the inverse system size, $2 \piup / L$. In this way, 
summations in equations~(\ref{time_series_density}) and 
(\ref{time_series_velocity}) automatically take care of minimum images of 
simulated particles.

\begin{figure}[!b]
\centerline{\includegraphics[scale=0.750]{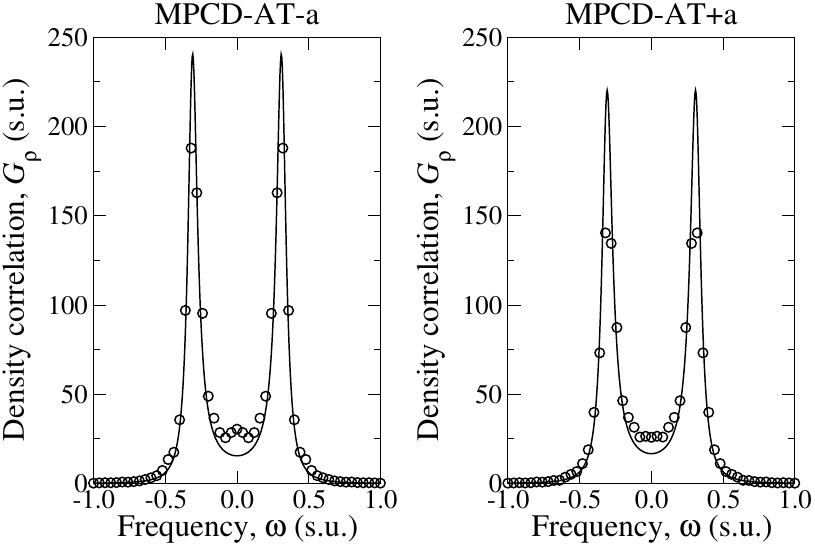}}
\caption{Density ACFs in MPCD-AT$\pm$a. Symbols represent results from numerical
implementations, while continuous curves are obtained from
equation~(\ref{density_acf}) complemented with the theoretical material 
parameters $c_{\textrm{T}}$ and $D_{\textrm{l}}$, expected for each MPCD-AT 
version. Reduced correlations are given by $G_{\rho}=C_{\rho}/A$.}
\label{figure_001}
\end{figure}

\begin{figure}[!b]
\centerline{\includegraphics[scale=0.750]{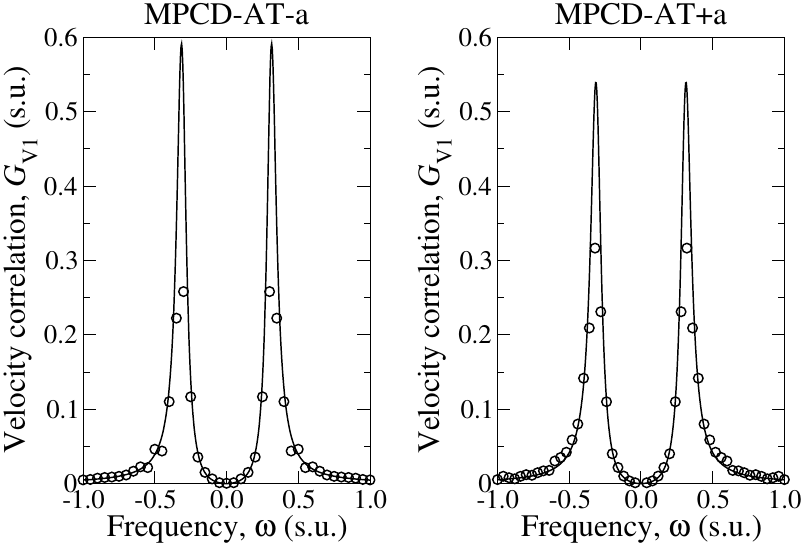}}
\caption{Same as in figure~\ref{figure_001} for reduced velocity ACFs 
$G_{\textrm{V1}}=
C_{\textrm{V1}}/A$.}
\label{figure_002}
\end{figure}

\begin{figure}[!t]
\centerline{\includegraphics[scale=0.750]{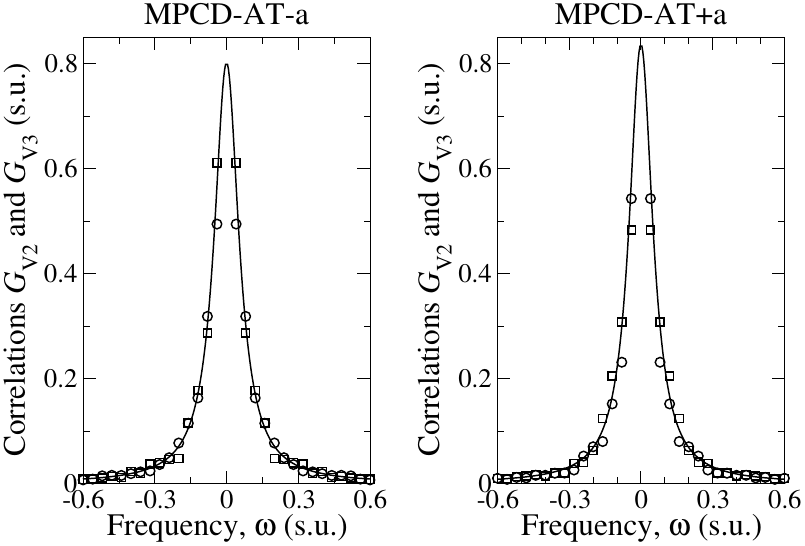}}
\caption{Velocity ACFs, $G_{\textrm{V2}}= C_{\textrm{V2}}/A$ and
$G_{\textrm{V3}}= C_{\textrm{V3}}/A$ in MPCD-AT$\pm$a. Symbols 
represent results from numerical implementations: $G_\text{V2}$ ($\bigcirc$); 
$G_\text{V3}$ ($\Box$). Continuous curves are obtained from 
equation~(\ref{velocity_trans_acf}) complemented with the 
theoretical viscosity, $\nu$, expected for each MPCD-AT version.}
\label{figure_003}
\end{figure}

Numerical experiments return ACF with the mathematical features summarised by 
equations~(\ref{density_acf})--(\ref{velocity_trans_acf}), which can be 
supplemented with an expression for viscosity coefficients, 
equations~(\ref{viscosity_kin_ma})--(\ref{viscosity_col_pa}), to verify
the correctness of the numerical implementation. 
Fixed units of time ($\Delta t=1.0$), energy ($k_{\textrm{B}} T = 1.0$), and 
mass ($m = 1.0$) were used during simulations. Units of length were derived 
from $a = \Delta t \sqrt{k_{\textrm{B}}T/m}$. Throughout the following, 
quantities will be reported in these simulation units (s.u.). Simulated systems 
had $L_{x} = L_{y} = L_{z} =L=20\,a$, and $\bar{N}^{\textrm{c}} = 20$. 
Scattering geometry was defined by the wave vector $\vec{k} = (2\piup/L,0,0)$.
In correlations obtained from MPCD-AT$\pm$a implementations, constant~$A$ 
in equations~(\ref{density_acf})--(\ref{velocity_trans_acf}) was considered an 
adjustable parameter. Reduced functions defined by
$G_{\rho} = C_{\rho}/A$, 
and 
$G_{\textrm{V}\alpha}
=C_{\textrm{V}\alpha}/A$, 
are compared against their theoretical counterparts obtained from 
equations~(\ref{density_acf})--(\ref{velocity_trans_acf}) and 
(\ref{viscosity_kin_ma})--(\ref{viscosity_col_pa}) in 
figures~\ref{figure_001}--\ref{figure_003}. They show the ability of MPCD to 
support hydrodynamic fluctuations that relax towards equilibrium as fluctuations 
do in a fluid kept in contact with a thermal bath. 
The good agreement found between numerical and theoretical results exhibits a 
very satisfactory understanding about MPCD that makes it a reliable method for 
simulations of soft condensed matter systems.

\section{MPCD for nematic phases}
\label{mpcd_nematic_section}

Liquid crystals are matter phases characterised by possessing a structural order 
lower than the one present in crystals, but larger than the corresponding to 
ordinary isotropic liquids~\cite{030}. They have been subject of continuous 
interest through decades since they exhibit anisotropic optical properties that 
can be manipulated by small energies of electromagnetic or surface-liquid 
interaction origin~\cite{031}. 
Modern LC technologies have promising applications in colloid manipulation, 
detection of biological agents, and medical diagnosing~\cite{031a,032,033,034}. 
Numerical simulations could play an important role in helping developers to test 
models of these emerging technologies~\cite{035}.

In nature, NLCs are constituted by anisotropic molecules of elongated or 
discotic types, commonly referred to as \emph{nematogens}. Accordingly, in order 
to generalise MPCD to nematics, simulated particles are supplied with an 
orientation degree of freedom $u_{i,\alpha} = u_{i,\alpha}(t)$, which is a unit
vector. Furthermore, MPCD rules described in
section~\ref{mpcd_algorithm_isotropic_section} are augmented to cover three
main aspects, namely: multi-particle collision events cause changes in 
$u_{i,\alpha}$; velocity gradients produce torques on $u_{i,\alpha}$; and 
reorientation induces flow (backflow effect). Implementation of these effects is 
described below.

\subsection{Orientation exchange}
\label{orientation_exchange_section}

Orientation exchange takes into account the interaction of nematogens with their 
surroundings in a coarse-grained fashion. More precisely, order in NLCs is 
measured through the quadrupolar moment of the orientation distribution, also 
referred to as the order parameter tensor. This quantity is estimated in MPCD-N 
at the cell level through
\begin{equation}
Q^{\textrm{c}}_{\alpha\beta}      
		       = \frac{1}{2 N^{\textrm{c}}} 
                         \sum_{j \in c} 
                         \left( 
                         3 u_{j,\alpha} u_{j,\beta} - \delta_{\alpha \beta}
                         \right),
\label{order_tensor}
\end{equation}
where $N_{\textrm{c}}$ is the number of particles in the cell after streaming 
and grid shift. For small-size nematic systems far away from the 
isotropic-nematic transition, the amount of the order in the cell can be quantified 
by the largest eigenvalue of $Q^{\textrm{c}}_{\alpha\beta}$, called the scalar 
order parameter, $S^{\textrm{c}}$. The unit eigenvector associated to 
$S^{\textrm{c}}$, $n^{\textrm{c}}_{\alpha}$, is called the director and 
represents the average molecular orientation. If nematogen $i$ is found in a 
cell with the order parameter $S^{\textrm{c}}$ and director 
$n^{\textrm{c}}_{\alpha}$, it is assumed that it interacts with nematogens 
within the same cell according to the (mean-field) Maier-Saupe potential 
energy~\cite{035a},
\begin{equation}
U_{\textrm{mf}} = -\frac{3}{2} U S^{\textrm{c}} 
		 \left(
		 u_{i,\alpha}u_{i,\beta}
                 n^{\textrm{c}}_{\alpha}n^{\textrm{c}}_{\beta} - 1
		 \right),
\label{mean_field_potential}
\end{equation}
where $U$ represents the strength of the mean-field interaction. Accordingly, an 
orientation collision operator can be proposed that updates orientations by 
sampling new vectors, $u_{i,\alpha}^{\prime}$, from the canonical distribution
$ P = Z^{-1} \exp\left[ -U_{\text{mf}}/(k_\text{B}T) \right]\sin\theta$, where 
$Z^{-1}$ is the normalisation constant and $\theta$ is the angle between 
$u_{i,\alpha}$ and $n^{\textrm{c}}_{\alpha}$.

\subsection{Reorientation by flow}
\label{reorientation_by_flow_section}

Flow effects on $u_{i,\alpha}$ are modelled as if nematogens were slender rods 
that experience torques induced by local velocity gradients. The velocity 
gradient tensor at a given collision cell is measured by considering the
difference between the centre of mass velocities of its neighbouring planes. Then,
nematogens are reoriented following the rule~\cite{038}
\begin{equation}
u_{i,\alpha} \left(t + \Delta t\right) 
= u_{i,\alpha}^{\prime} 
+ \chi_{\text{HI}} 
  \left[
  \Omega^{\textrm{c}}_{\alpha\beta}u_{i,\beta}^{\prime}
+\lambda \left(D^{\textrm{c}}_{\alpha\beta} u_{i,\beta}^{\prime} 
-D^{\textrm{c}}_{\alpha^{\prime}\beta^{\prime}}u_{i, \beta^{\prime}}^{\prime} 
                                      u_{i,\alpha^{\prime}}^{\prime} 
                                      u_{i,\alpha}^{\prime}\right)
  \right],
\label{shear_flow_alignment}
\end{equation}
where $\Omega^{\textrm{c}}_{\alpha \beta}$ and 
$D^{\textrm{c}}_{\alpha \beta}$ stand for the antisymmetric and symmetric
parts of $\partial v_{\alpha}/\partial x_{\beta}$, 
respectively.  Equation~(\ref{shear_flow_alignment}) introduces quantities 
$\lambda$ and $\chi_{\text{HI}}$. The former is the tumbling parameter. 
It determines the tendency of $u_{i,\alpha}$ to be aligned by shear at a 
stationary angle ($\lambda>1$), or to continuously rotate in a direction 
consistent with the vorticity of the flow ($\lambda < 1$). During simulations, 
$\lambda$ can be fixed to produce the desired dynamic effect~\cite{016}. 
Parameter $\chi_{\text{HI}}$ is an auxiliary one. It is initialised within the 
range $[0,1]$ to control the overall flow-orientation coupling. By setting 
$\chi_{\text{HI}} = 0$, such a coupling vanishes, while the maximum reorientation by 
flow is simulated for $\chi_{\text{HI}}=1$.

\subsection{Backflow}
\label{backflow_section}

During the orientation update stage, cells receive an effective angular velocity
\begin{equation}
w^{\textrm{c}}_{\alpha}
=-\frac{1}{\Delta t}\sum_{j \in c} 
\varepsilon_{\alpha \alpha^{\prime} \beta^{\prime}} 
u_{j,\alpha^{\prime}}(t+\Delta t)
u_{j,\beta^{\prime}}(t).
\label{angular_velocity}
\end{equation}
Under the assumption of over-damped rotational dynamics with viscosity 
$\gamma_{\text{R}}$, this implies a net torque, 
$\Gamma^{\textrm{c}}_{\alpha} = -\gamma_\text{R} w^{\textrm{c}}_{\alpha}$, and the 
production of an angular momentum in the cell 
$\Delta{\mathcal L}^{\textrm{c}}_{\alpha}=\Gamma^{\textrm{c}}_{\alpha}\Delta t$.

Backflow is taken into account by adding this angular momentum contribution to
$\Delta L^{\textrm{c}}_{\alpha}$ in the MPCD-AT$+$a collision rule, 
equation~(\ref{collision_pa}). 

\section{Fluctuating nematodynamics}
\label{fluctuating_nematodynamics_section}

In MPCD-N, in addition to density and velocity fluctuations discussed in 
section~\ref{hydrodynamic_fluctuations_mpcd_section}, there occur spontaneous 
changes in the orientation field. Orientation fluctuations are defined through
\begin{equation}
n_{\alpha}\left(\vec{r},t\right) = n^{\textrm{eq}}_{\alpha} 
                                 + \delta n_{\alpha}\left(\vec{r},t\right),
\label{director_fluctuation_def}
\end{equation}
where $n^{\textrm{eq}}_{\alpha}$ is the global director field, which is assumed 
to be fixed in space. Thermal director fluctuations can be analysed by extending 
the linearized scheme of section~\ref{hydrodynamic_fluctuations_mpcd_section}. 
Specifically, following general hydrodynamic models of NLCs~\cite{038a}, the 
relaxation equation for the director field is proposed to have the form
\begin{equation}
\left( \frac{\partial }{\partial t} 
      + V_{\beta} \frac{\partial}{\partial x_{\beta}}
\right) 
n_{\alpha} = Y_{\alpha} + \Upsilon_{\alpha}\, ,
\label{director_relaxation}
\end{equation}
where $Y_{\alpha}$ is sometimes termed a \emph{quasi-current} due to the fact 
that the director is not a conserved quantity and, consequently, the integral of 
$Y_{\alpha}$ over a surface cannot be a flux~\cite{038a}. In 
equation~(\ref{director_relaxation}), $\Upsilon_{\alpha}$ is a stochastic 
contribution to $Y_{\alpha}$, which can be assumed to be a Gaussian-Markov 
process with zero mean, strength $\gamma$, and FDT
\begin{equation}
\left\langle \Upsilon_{\alpha} \left(\vec{r}^{\,\prime},t^{\prime}\right)  
             \Upsilon_{\beta}  \left(\vec{r},t\right)  
\right\rangle = 2 k_\text{B} T \gamma 
                \left(
                \delta_{\alpha\beta} - n_{\alpha} n_{\beta}
                \right) 
                \delta\left(\vec{r}^{\,\prime} -\vec{r}\right)
                \delta\left(t^{\prime}-t\right),
\label{quasi_current_fdt}
\end{equation}
where matrix $\delta_{\alpha \beta} - n_{\alpha} n_{\beta}$ projects vectors 
on a direction perpendicular to $n_{\alpha}$. This permits to satisfy the 
condition $n_{\alpha} \delta n_{\alpha} = 0$, imposed by the normalisation
of $n_{\alpha}$.

In MPCD-N, time evolution of orientations is promoted by shear, 
equation~(\ref{shear_flow_alignment}). Furthermore, streaming and collision 
dynamics could be anticipated to give rise to orientation diffusion, then 
$Y_{\alpha}$ is written as
\begin{equation}
Y_{\alpha}    =  -D_{n} \left( 
                          \delta_{\alpha \beta} - n_{\alpha} n_{\beta} 
                          \right) 
                          \frac{\partial }{\partial x_{\beta^{\prime}}} 
                          \frac{\partial}{\partial x_{\beta^{\prime}}} 
                          n_{\beta} 
             + \chi_{\text{HI}} 
               \left[ 
	       \Omega_{\alpha \beta} n_{\beta}
             + \lambda \left(
		     D_{\alpha \beta} n_{\beta}
		    -D_{\alpha^{\prime}\beta^{\prime}} 
                     n_{\beta^{\prime}}n_{\alpha^{\prime}}n_{\alpha}
		     \right)
	     \right],
\label{quasi_current_exp}
\end{equation}
where the orientation diffusion coefficient is $D_{n}$ and 
$\Omega_{\alpha\beta}$ and $D_{\alpha \beta}$ are the antisymmetric and symmetric
parts of $\partial V_{\alpha}/\partial x_{\beta}$. 

Dynamic equations for director fluctuations can be obtained by replacing 
equations~(\ref{velocity_fluctuation}) and (\ref{director_fluctuation_def}) 
into equations~(\ref{director_relaxation}) and (\ref{quasi_current_exp}), and
retaining only linear contributions. 
In a first approximation, mass and momentum transport can be assumed to be still 
given by equations~(\ref{continuity}) and (\ref{navier_stokes}), thus 
neglecting the effects of backflow~\cite{039}. In 
section~\ref{measurements_nem_section}, the validity of this assumption will be 
explored. It will be shown to be a very satisfactory approximation
for those values $\gamma_{\textrm{R}}$ used up to now in previous implementations
of MPCD-N.
As in the case of simple fluids, the resulting system is more conveniently 
studied in Fourier space. In addition, separation of transverse and longitudinal
variables is worth recommending. Taking into account nematic symmetry, basis
vectors are defined by $\hat{e}_{1} = \vec{k}/k$, 
$\hat{e}_{2} = (\hat{n}^{\textrm{eq}} \times \hat{e}_{1})/\sin\theta^{\textrm{eq}}$, and 
$\hat{e}_{3} = \hat{e}_{1} \times \hat{e}_{2}$, where $\theta^{\textrm{eq}}$ is the angle 
between $\hat{n}^{\textrm{eq}}$ and $\vec{k}$.
For concreteness, the scattering geometry defined by $\vec{k} = (k,0,0)$, and 
$\hat{n}^{\textrm{eq}} = (0,0,1)$ will be considered hereafter. Then, the following system
of independent equations is obtained for fluctuating fields in MPCD-N,
\begin{eqnarray}
\left( 
       \begin{matrix}
       -\ri \omega + D_{n} k^{2} & 0 \\
       0 & -\ri \omega + \nu k^{2} 
       \end{matrix}
\right) 
\cdot
\left(
       \begin{matrix}
          \delta \tilde{n}_{2} \\
          \delta \tilde{V}_{2}
       \end{matrix}
\right) 
=
\left(
       \begin{matrix}
          \tilde{\Upsilon}_{2} \\
       \ri k\rho_{0}^{-1}\tilde{\Sigma}_{12}
       \end{matrix}
\right) ,
\label{transverse_nem}
\end{eqnarray}
\begin{eqnarray}
\left(\begin{matrix}
        -\ri \omega                    &  -\ri k \rho_{0} &  \\
	-\ri k c_{T}^{2} \rho_{0}^{-1} & -\ri \omega  + D_{\textrm{l}} k^{2}
      \end{matrix}
\right) \cdot 
\left(\begin{matrix}
      \delta\tilde{\rho} \\
      \delta\tilde{V}_{1} 
      \end{matrix}
\right) =
        \frac{\ri k}{\rho_{0}} 
\left(\begin{matrix}
      0 \\
      \tilde{\Sigma}_{11}
      \end{matrix}
\right) ,
\label{longitudinal_nem2}
\end{eqnarray}
and
\begin{eqnarray}
\left( 
       \begin{matrix}
       -\ri \omega + D_{n} k^{2} & \ri k \chi_{\text{HI}} (\lambda -1)/2 \\
       0 & -\ri \omega + \nu k^{2} 
       \end{matrix}
\right) 
\cdot
\left(
       \begin{matrix}
          \delta \tilde{n}_{1} \\
          \delta \tilde{V}_{3}
       \end{matrix}
\right) 
=
\left(
       \begin{matrix}
          \tilde{\Upsilon}_{1} \\
       \ri k\rho_{0}^{-1}\tilde{\Sigma}_{13}
       \end{matrix}
\right) .
\label{longitudinal_nem1}
\end{eqnarray}

It can be observed from 
equations~(\ref{transverse_nem})--(\ref{longitudinal_nem1}) that in the present
nematodynamic model, density and velocity fluctuations are not perturbed
by orientation fluctuations though fluctuations $\delta \tilde{V}_{3}$ affect 
the dynamics of $\delta \tilde{n}_{1}$. Consequently, it is expected that
density and velocity ACFs in MPCD and MPCD-N will be the same. 

Orientation fluctuations can be easily solved from 
equations~(\ref{transverse_nem}) and (\ref{longitudinal_nem1}). Such a solution, 
together with the FDT in Fourier space,
\begin{eqnarray}
\big\langle \tilde{\Upsilon}_{\alpha}    \big( \vec{k},\omega \big) 
	     \tilde{\Upsilon}^{*}_{\beta}\big( \vec{k},\omega \big)\big\rangle
= 2 (2\piup)^{4} \gamma k_\text{B} T \left(\delta_{\alpha\beta}
	                           -n_{\alpha} n_{\beta}\right) 
                              \delta\big( \vec{k}^{\prime} +\vec{k}\big) 
	                      \delta\left(\omega^{\prime}+\omega\right),
\label{longitudinal_nem}
\end{eqnarray}
can be used to obtain the following orientation ACFs
\begin{eqnarray}
C_{\textrm{n2}} \big(\vec{k},\omega \big)
       = \big\langle 
         \delta \tilde{n}_{2}\big( \vec{k},\omega \big) 
         \delta \tilde{n}_{2}^{*}\big( \vec{k},\omega \big) 
         \big\rangle
       = A \frac{ \gamma }{\omega^{2} + D_{\textrm{n}}^{2}k^{4}} \,,
\label{trans_dir_acf}
\end{eqnarray}
\begin{eqnarray}
C_{\textrm{n1}} \big(\vec{k},\omega \big)
     =  \big\langle 
         \delta \tilde{n}_{1}\big( \vec{k},\omega \big) 
         \delta \tilde{n}_{1}^{*}\big( \vec{k},\omega \big) 
         \big\rangle 
     = A\left[
	    \frac{\gamma}{\omega^{2} + D_{\textrm{n}}^{2}k^{4}} 
	    +\frac{\chi_{\text{HI}}^{2}(\lambda-1)^{2}}{4 \rho_{0}}
	     \frac{\nu k^{4}}{\left(\omega^{2} + D_{\textrm{n}}^{2}k^{4}\right)
                              \left(\omega^{2} +   \nu^{2}k^{4}\right)}
          \right] .
\label{long_dir_acf}
\end{eqnarray}

Consequently, $C_{\textrm{n2}}$ is not expected to be 
modified by velocity-orientation coupling, while the intensity of 
$C_{\textrm{n1}}$ is anticipated to increase with a term
proportional to $\chi_{\text{HI}}^{2} (\lambda-1)^{2}$.

Another correlation function that could bring information about the correct
coupling between velocity and orientation fluctuations is the cross correlation
function, $C_{\textrm{n1,V3}}$, given by
\begin{equation}
C_{\textrm{n1,V3}} \big(\vec{k},\omega\big) =
\big\langle
         \delta \tilde{n}_{1}\big(\vec{k},\omega\big)
         \delta \tilde{V}_{3}\big(\vec{k},\omega\big)
\big\rangle 
=
A \frac{\chi_{\text{HI}} (\lambda - 1)}{2 \rho_{0}} 
\frac{\nu k^{3} (\ri \omega + D_{n} k^{2})}{\left(\omega^{2} + D_{n}^{2}k^{4}\right)
        \left(\omega^{2} +   \nu^{2}k^{4}\right)}.
\label{cross_cf}
\end{equation}

\subsection{Measurements}
\label{measurements_nem_section}

Measurements of correlations $C_{\textrm{n1}}$, $C_{\textrm{n2}}$, and 
$C_{\textrm{n1,V3}}$ can be carried out by introducing the time-series of 
spatial Fourier transform of orientation fluctuations
\begin{equation}
\delta\bar{n}_{\alpha} \big(\vec{k}, t_{l}\big) 
= \sum_{j=1}^{N} \left(
		 u_{j,\alpha} - n^{\textrm{eq}}_{\alpha}
		 \right)
		 \exp\big[-\ri \vec{k} \cdot \vec{r}_{j}(t_{l}) \big].
\label{time_series_dir}
\end{equation}

Then, the same procedure used in section~\ref{measurements_section} to measure 
density and velocity ACFs can be followed. It is worth 
stressing that in order to make these measurements compatible with the use of 
PBC and with the model yielding 
equations~(\ref{trans_dir_acf})--(\ref{cross_cf}), it is necessary to ensure 
that global director will remain fixed during the computation stage. This can be 
done by introducing an additional step in the MPCD-N algorithm where all 
nematogens are rotated by the same angle in order to align $\hat{n}^{\textrm{eq}}$ 
towards the $z$~\cite{040}. This direction is 
considered as the $\hat{e}_{3}$ direction, while $x$ and $y$ are considered to 
coincide with $\hat{e}_{1}$ and $\hat{e}_{2}$, respectively.

Numerical experiments were conducted using the same simulation parameters 
$\Delta t$, $k_{\textrm{B}} T$, $m$, $\bar{N}^{\textrm{c}}$, $a$, and $L$, given 
in section~\ref{measurements_section}. Nematic order was simulated using 
$U=20k_\text{B}T$. For this mean-field strength, it has been shown that the average 
scalar order parameter produced in collision cells, 
$\bar{S}^{\textrm{c}}=0.947$, is in agreement with predictions of 
self-consistent models of NLCs~\cite{039}. Two
main features of hydrodynamic correlation functions in MPCD-N will be explored.
Namely, the effects of backflow controlled by the parameter 
$\gamma_{\textrm{R}}$; and the effects of velocity-orientation coupling 
controlled by parameters $\chi_{\text{HI}}$ and $\lambda$.

\subsubsection{Backflow effects}
\label{backflow_effects_section}

\begin{figure}[!t]
\centerline{\includegraphics[scale=0.750]{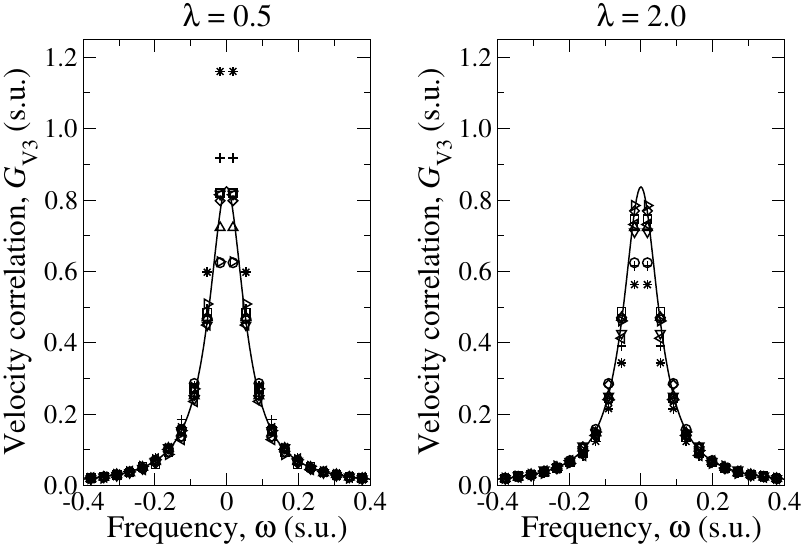}}
\caption{Reduced velocity ACF $G_{\textrm{V3}}$ in MPCD-N. Continuous curves are 
obtained from equation~(\ref{velocity_trans_acf}) and represent the expected
behaviour for MPCD-AT$+$a. Symbols correspond to numerical results obtained from
variations of the tumbling parameter, $\lambda$; the flow-director coupling 
parameter, $\chi_{\textrm{HI}}$; and the rotation viscosity, 
$\gamma_{\textrm{R}}$. The symbology used is defined as follows: 
$\gamma_{\textrm{R}} = 0.01$, $\chi_{\textrm{HI}} = 0.0$ ($\bigcirc$);
$\gamma_{\textrm{R}} = 0.01$, $\chi_{\textrm{HI}} = 0.5$~($\Box$);
$\gamma_{\textrm{R}} = 0.01$, $\chi_{\textrm{HI}} = 1.0$ ($\Diamond$);
$\gamma_{\textrm{R}} = 0.10$, $\chi_{\textrm{HI}} = 0.0$ ($\triangle$);
$\gamma_{\textrm{R}} = 0.10$, $\chi_{\textrm{HI}} = 0.5$ ($\bigtriangledown$);
$\gamma_{\textrm{R}} = 0.10$, $\chi_{\textrm{HI}} = 1.0$~($\lhd$);
$\gamma_{\textrm{R}} = 1.00$, $\chi_{\textrm{HI}} = 0.0$ ($\rhd$);
$\gamma_{\textrm{R}} = 1.00$, $\chi_{\textrm{HI}} = 0.5$ ($+$);
$\gamma_{\textrm{R}} = 1.00$, $\chi_{\textrm{HI}} = 1.0$~($\ast$).}
\label{figure_004}
\end{figure}

With the purpose of analysing the extent at which the backflow perturbs 
hydrodynamic fluctuations, $\gamma_{\textrm{R}}$ was varied over the range 
$\left[0.01,1.0\right]$, which extends by one order of magnitude the values 
considered before the present contribution. 
Eighteen experiments were conducted using combinations between parameters 
$\lambda = 0.5, 2.0$; $\chi_{\textrm{HI}} = 0.0, 0.5, 1.0$, and 
$\gamma_{\text{R}} = 0.01, 0.1, 1.0$. Correlation functions 
$C_{\rho}$ and $C_{\textrm{V}\alpha}$ were measured and compared with those
obtained from simple MPCD-AT$+$a. No systematic differences were observed
between correlations $C_{\rho}$, $C_{\textrm{V1}}$ and $C_{\textrm{V2}}$ in 
MPCD-N and those measured in MPCD-AT$+$a, in good agreement with 
equations~(\ref{transverse_nem}) and (\ref{longitudinal_nem2}). A similar 
conclusion applies for function $C_\text{V3}$ for small and moderate values of the
product $\gamma_{\textrm{R}} \chi_{\textrm{HI}}$, namely, for 
$\gamma_{\textrm{R}} \chi_{\textrm{HI}} < 0.5$. Nevertheless, backflow effects
become noticeable, increasing the intensity of $C_{\textrm{V3}}$ for $\lambda =0.5$,
and decreasing it for $\lambda = 2.0$. This is illustrated in 
figure~\ref{figure_004}, where the results for $C_\text{V3}$ obtained for the whole
experimental setup are presented. 
Therefore, it can be concluded that if $\gamma_{\textrm{R}} \leqslant 0.1$, as it has 
been used in references~\cite{016,039}, hydrodynamic fluctuations of flow and 
density in MPCD-N can be considered to be independent of orientation 
fluctuations.

\subsubsection{Velocity-orientation coupling}
\label{velocity_orientation_coupling_section}

In order to analyse the dynamics of director fluctuations, reduced 
correlations are introduced as follows:
\begin{equation}
G_{\textrm{n2}} \big(\vec{k},\omega\big)
= \frac{C_{\textrm{n2}}\big(\vec{k},\omega\big)}{A} 
= \frac{\gamma}{\omega^{2} + D_{\textrm{n}}^{2} k^{4}} \,,
\label{reduced_n_acf_trans}
\end{equation}
\begin{equation}
G_{\textrm{n1}}\big(\vec{k},\omega\big) 
= \frac{C_{\textrm{n1}}\big(\vec{k},\omega\big)}{A} 
= \frac{\gamma}{\omega^{2} + D_{\textrm{n}}^{2} k^{4}} 
	+\frac{\chi_{\text{HI}}^{2}(\lambda-1)^{2}}{4 \rho^{\textrm{eq}}}
	     \frac{\nu k^{4}}{\left(\omega^{2} + D_{\textrm{n}}^{2}k^{4}\right)
                              \left(\omega^{2} +   \nu^{2}k^{4}\right)}\,,
\label{reduced_n_acf_longi}
\end{equation}
and
\begin{equation}
G_{\text{n1,V3}} \big(\vec{k},\omega\big) 
	  = \frac{C_{\text{n1,V3}}\big(\vec{k},\omega\big)}{A} 
	  = \frac{\chi_{\text{HI}} (\lambda - 1)}{2 \rho^{\textrm{eq}}} 
	    \frac{\nu k^{3} ( -\omega + \ri D_{\textrm{n}} k^{2})}
	    {\left(\omega^{2} + D_{\textrm{n}}^{2}k^{4}\right)
            \left(\omega^{2} +   \nu^{2}k^{4}\right)}  .
\label{reduced_n_acf_cross}
\end{equation}

Numerical results are normalised using the value of $A$ obtained in simulations
of isotropic phases.

\begin{figure}[!t]
\centerline{\includegraphics[scale=0.750]{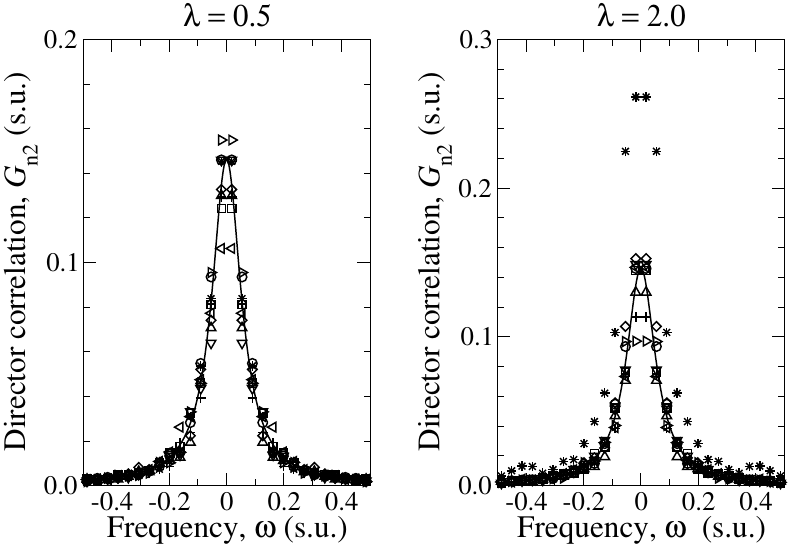}}
\caption{Same as in figure~\ref{figure_004} for reduced transverse director 
ACF $G_{\text{n2}}(\vec{k},\omega)$. Continuous curves are obtained 
from equation~(\ref{reduced_n_acf_trans}) using empirical values for
$A$, $\gamma$, and $D_{\textrm{n}}$. Symbols are numerical results. Symbology
is the same as the one introduced in figure~\ref{figure_004}. Discrepancy 
between numerical experiments and theory is observed for the case 
$\lambda=2.0$ at $\gamma_{\text{R}}=2.0$, and $\chi_{\text{HI}} = 1.0$ ($\ast$).}
\label{figure_005}
\end{figure}

\begin{figure}[!t]
\centerline{\includegraphics[scale=0.675]{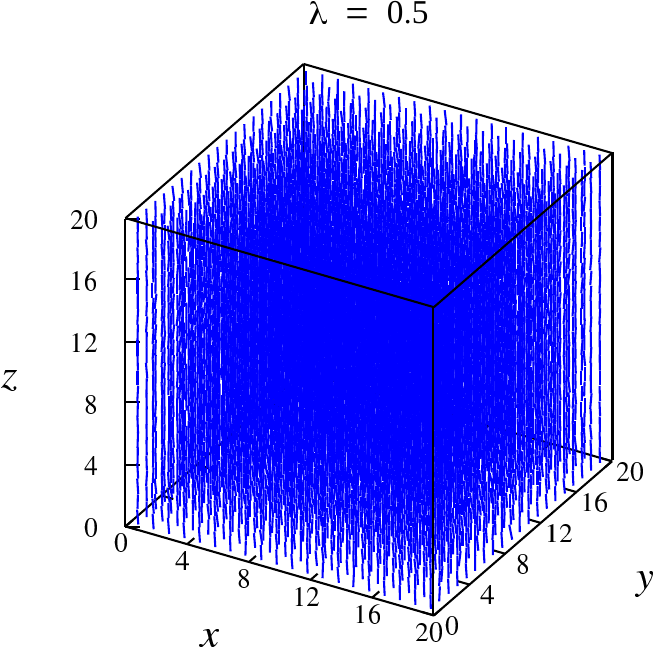} 
            \includegraphics[scale=0.675]{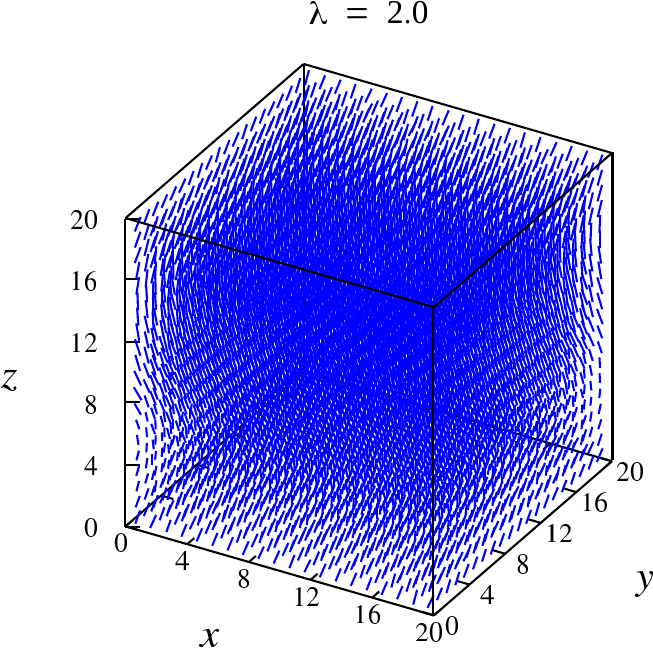}}
\caption{(Colour online) Order in simulated systems at maximum velocity-orientation coupling 
($\chi_{\text{HI}} = 1$) and backflow ($\gamma_{\textrm{R}}$), for two different 
values of the tumbling parameter, $\lambda$. Short lines are used to represent 
the director at collision cells, $n^{\textrm{c}}_{\alpha}$. Snapshots represent 
the state of the system in the last simulation step. At $\lambda = 2.0$, velocity 
fluctuations diminish the global order.}
\label{figure_006}
\end{figure}

It is worth stressing that, up to the present, there are no analytical 
treatments of MPCD-N from which values of $\gamma$ and $D_{\textrm{n}}$ 
could be inferred in terms of simulation parameters, $\bar{N}^{\textrm{c}}$, 
$\Delta t$, $m$, etc. Consequently, the subsequent discussion is conducted in a 
semiempirical form, using values for $\gamma$ and $D_{\textrm{n}}$ derived from 
the numerical experiments.
Correlations $G_{\textrm{n2}}$ and $G_{\textrm{n1}}$ showed the Lorentzian shape 
predicted from equations~(\ref{reduced_n_acf_trans}) and 
(\ref{reduced_n_acf_longi}) with estimates, in s.u., 
$\gamma_{\text{R}}=5.5\times 10^{-4}$ and $D_{\text{n}} = 0.615$. 
$G_{\textrm{n2}}$ was found to be independent of 
simulation parameters, as expected from equation~(\ref{reduced_n_acf_trans}),
except for simulations conducted at $\gamma_{\text{R}} = 1.0$, $\lambda = 2.0$, 
and $\chi_{\textrm{HI}}= 1.0$. In this case, $G_{\textrm{n2}}$ is more intense 
and exhibits two small lateral peaks located at $\pm q(k)$, indicating an 
apparent coupling with propagating modes. These features are illustrated in 
figure~\ref{figure_005}. This behaviour can be explained by noting that 
velocity fluctuations play a destructive role on the orientation order, since they 
increase the strength of director fluctuations~\cite{016}. At large values of 
$\lambda$ and $\chi_{\textrm{HI}}$, this effect could be large enough to produce
a reduction of global order in the system. Specifically, for $\lambda = 2.0$ and
$\chi_{\text{HI}}=1.0$, the average global order in the sample was found to be 
$S = 0.74$, while in all remaining cases it was found to be close to 
$S = 0.94$. This is schematically illustrated in figure~\ref{figure_006} where 
snapshots of systems simulated at 
$\{\lambda = 0.5,\gamma_{\textrm{R}}=1.0,\chi_{\text{HI}} = 1.0\}$ and 
$\{\lambda = 2.0,\gamma_{\textrm{R}}=1.0,\chi_{\text{HI}} = 1.0\}$ are shown. 
Large director gradients can be observed in the latter case, for which it could 
be expected that significant changes of the global director will be produced 
between simulation steps. Then, the approximation of fixed 
$\hat{n}^{\textrm{eq}}$, from which separation of longitudinal and transverse 
variables is constructed, becomes weaker. This implies that in the
case $\{\lambda = 2.0,\gamma_{\textrm{R}}=1.0,\chi_{\text{HI}} = 1.0\}$, 
fluctuation $\delta \tilde{n}_{y}$ that is used to estimate $G_{\textrm{n2}}$,
could be coupled with the propagating velocity components, a situation not expected 
for the transverse variable $\delta \tilde{n}_{2}$.

\begin{figure}[!t]
\centerline{\includegraphics[scale=0.750]{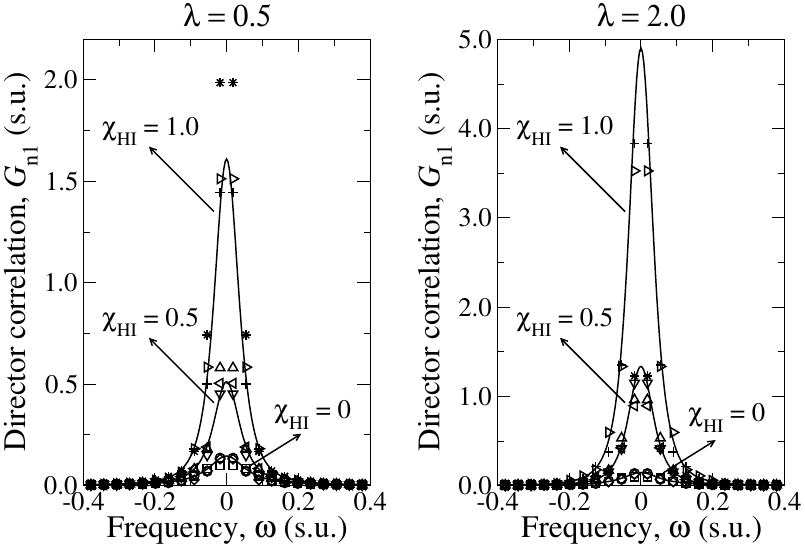}}
\caption{Reduced longitudinal director ACF $G_{\text{n1}}$ in MPCD-N. Symbols
have the same meaning as in figure~\ref{figure_004}. Curves correspond to 
predictions of the linearized model summarised by 
equation~(\ref{reduced_n_acf_longi}). For both, $\lambda = 0.5$ and 
$\lambda = 2.0$, the theoretical model does not fit numerical results at 
$\gamma_{\textrm{R}} =1.0$, and $\chi_{\textrm{HI}} =1.0$~($\ast$).}
\label{figure_007}
\end{figure}

\begin{figure}[!t]
\centerline{\includegraphics[scale=0.750]{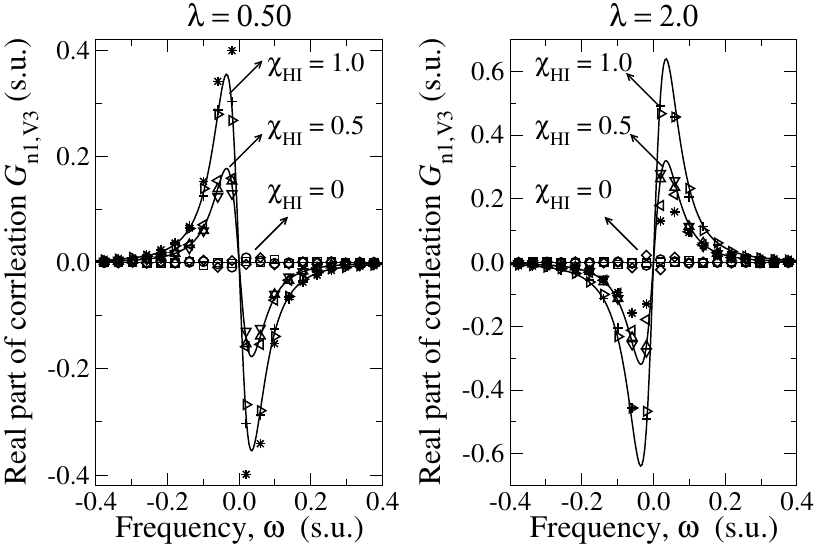}}
\caption{Real part of the reduced cross correlation $G_{\textrm{n1,V3}}$. 
Theoretical curves are obtained from equation~(\ref{reduced_n_acf_cross}). 
Symbols have the same meaning as in figure~\ref{figure_004}. Numerical
and theoretical predictions coincide except at cases 
$\gamma_{\textrm{R}} =1.0$, and $\chi_{\textrm{HI}} =1.0$ ($\ast$).} 
\label{figure_008}
\end{figure}

Correlation $G_{\textrm{n1}}$ was found to exhibit a
systematic dependence on the product $(\lambda -1)^{2}\chi_{\text{HI}}^{2}$.
Results are in good agreement with equation~(\ref{reduced_n_acf_longi}), as it
can be observed in figure~\ref{figure_007}. Analogous results were found in
the case of the cross correlation $G_{\textrm{n1,V3}}$,
whose real part obtained from equation~(\ref{reduced_n_acf_cross}) is illustrated 
in figure~\ref{figure_008} and is compared with that obtained from numerical 
experiments. Thus, it is shown that the coupling between orientation and velocity 
fields predicted from the linearized model of nematic phases is well reproduced 
by MPCD-N. Again, systems simulated at $\gamma_{\textrm{R}}= 1.0$, and 
$\chi_{\text{HI}} = 1.0$, seem to deviate from this rule. In the case 
$\lambda = 0.5$, $G_{\textrm{n1}}$ and $G_{\textrm{n1,V3}}$ had a larger height 
than the one predicted by the linearized model. On the contrary, for 
$\lambda = 2.0$, both $G_{\textrm{n1}}$ and $G_{\textrm{n1,V3}}$ decreased their 
maximum  height with respect to the value anticipated by the theoretical model. 
This discrepancy is related with the behaviour of the velocity ACF
$C_{\textrm{V3}}$ already presented in figure~\ref{figure_004}. To be specific,
the increment (decrement) of $C_{\textrm{V3}}$ for $\lambda=0.5$ ($\lambda=2.0$)
causes a corresponding increment (decrement) of $G_{\textrm{n1}}$ and 
$G_{\textrm{n1,V3}}$ through the hydrodynamic coupling in 
equation~(\ref{longitudinal_nem1}).

\section{Conclusions}
\label{conclusions_section}

The capacity of MPCD-AT for reproducing collective phenomena of the hydrodynamic 
type has been analysed by exploring its spectra of thermal fluctuations in 
implementations for simple fluids and LCs. First, it was illustrated that 
relaxation of spontaneous fluctuations in simple fluids simulated under 
MPCD-AT$\pm$a rules can be described by linearized hydrodynamics. Analytical 
expressions of viscosity coefficients have been used for comparison purposes, 
in order to exhibit a good agreement existing between theoretical models and 
numerical implementations of MPCD-AT$\pm$a.

Afterwards, an extension of MPCD towards nematic phases has been presented. This 
implementation was based on the original one introduced in reference~\cite{016}. 
Considering the fundamental physical and mathematical aspects involved in the
study of LCs, as well as the technological relevance of these materials, such 
an extension could be very fruitful in the near future~\cite{035}. The analysis of 
correlation functions in MPCD-N was performed in a semiempirical form, since 
there are no explicit expressions for this method relating transport 
coefficients and simulation parameters. Numerical experiments showed that 
hydrodynamic fluctuations in MPCD-N can be described by a simplified linearized 
model where viscosity and elastic effects are isotropic. In addition, if 
simulation parameters are kept within the ranges of the original MPCD-N 
proposal~\cite{016}, velocity and orientation fluctuations are only one-way 
coupled. Namely, while velocity fluctuations affect the orientation fluctuations, 
backflow can be neglected in the linearized regime. Nonetheless, for large 
values of the parameters that couple the flow and orientation fields, 
$\gamma_{\textrm{R}}$ and $\chi_{\textrm{HI}}$, it was demonstrated that 
the velocity-orientation interaction can be strong enough to cause changes in the 
global orientation state. Though this effect has been already discussed in 
reference~\cite{016}, in the context of isotropic-nematic phase transition in 
MPCD-N, its implications on the measurements of hydrodynamic correlations were 
exhibited here for the first time. In this respect, it was concluded that large 
shear-induced reorientations could make the linearized model incompatible with 
correlation measurements. 

Numerical experiments were carried out in small-sized systems. Ongoing tests
are in progress for systems with larger sizes. Their purpose is to assess
the importance of size-dependence on the previously mentioned effects. Similarly, 
activities are currently under way that have the purpose of bringing MPCD-N
properties closer to those observed in real nematics. They include the 
introduction of director-dependent collision rules to simulate systems with 
anisotropic viscosity and elasticity, as well as the use of adapted mean field 
potentials in order to simulate smectic phases. 

\section*{Acknowledgements}

Author thanks La Salle University Mexico for final support under grant 
NEC-$08/18$.


\ukrainianpart

\title{Гідродинамічні кореляції в ізотропних плинах і рідких кристалах, модельовані з допомогою динаміки багаточастинкових зіткнень}

\author{Х. Хіяр}
\address{Інженерна школа, університет Ла Саль Мексики,  06140, Мехіко, Мексика}

\makeukrtitle

\begin{abstract}
 Динаміка багаточастинкових зіткнень --- це зручна числова методика для моделювання плинів на мезоскопічних масштабах. Вона дає змогу розглядати молекулярні деталі в огрублений спосіб і відтворювати гідродинамічні явища.  У даній статті, застосування динаміки багаточастинкових зіткнень  до ізотропних плинів проаналізовано за так званою термостатованою схемою Андерсена, конкретного алгоритма для систем у канонічному ансамблі. Цей метод виявляє гідродинамічні флуктуації, які спонтанно релаксують до рівноваги. Цей релаксаційний процес можна описати з допомогою лінеаризованої теорії та використати для обчислення коефіцієнтів переносу в системі. 
 Також розглянуто розширення алгоритму на нематичні рідкі кристали. Показано, що теплові коливання середньої орієнтації молекули можна описати за допомогою розширеної лінеаризованої схеми.  Флуктуації потоку індукують орієнтаційні флуктуації. Проте, орієнтаційні зміни мають помітний вплив на кореляційні функції швидкостей тільки, коли параметри моделювання перевищують їх значення у порівнянні зі значеннями, що використовуються у попередніх застосуваннях методу. В протилежному випадку, даний потік можна вважати незалежним від поля орієнтації.

\keywords  мультимасштабний метод моделювання, частинково базоване моделювання плинів, теплові флуктуації, нематичні рідкі кристали, структура рідин і рідких кристалів
\end{abstract}
\end{document}